\documentclass[sigconf]{acmart}

\copyrightyear{2021}
\acmYear{2021}
\setcopyright{rightsretained}
\acmConference[CAIN'22]{1st Conference on AI Engineering - Software Engineering for AI}{May 16--24, 2021}{Pittsburgh, PA, USA}
\acmBooktitle{1st Conference on AI Engineering - Software Engineering for AI (CAIN'22), May 16--24, 2021, Pittsburgh, PA, USA}
\acmDOI{10.1145/3522664.3528621}
\acmISBN{978-1-4503-9275-4/21/05}

\usepackage{tabularx}
\usepackage{hyperref}
\hypersetup{
  colorlinks=true,
  linkcolor=blue,
  filecolor=magenta,
  urlcolor=cyan
}
\usepackage{float}
\usepackage{graphicx}
\usepackage{listings}
\usepackage{caption}
\usepackage{subcaption}
\usepackage{calculator}
\newcommand{\myMultiply}[2]{%
    \MULTIPLY{#1}{#2}{\myV}
    \GLOBALCOPY{\myV}{\myV}}
\newcommand{\mybar}[2]{\myMultiply{#2}{1.0}\textcolor{#1}{\rule{\myV pt}{1ex}}}
\newcommand{\catbar}[1]{\mybar{teal!60}{#1}}
\newcommand{\subcatbar}[1]{\mybar{purple}{#1}}

\author{Arumoy Shome}
\affiliation{
  \institution{Delft University of Technology}
  \country{Netherlands}
}
\email{a.shome@tudelft.nl}

\author{Lu{\'\i}s Cruz}
\affiliation{
  \institution{Delft University of Technology}
  \country{Netherlands}
}
\email{l.cruz@tudelft.nl}

\author{Arie van Deursen}
\affiliation{
  \institution{Delft University of Technology}
  \country{Netherlands}
}
\email{arie.vandeursen@tudelft.nl}

\date{\textit{[2021-12-02 Thu]}}
\title{Data Smells in Public Datasets}
\begin{document}

\begin{abstract}
\emph{\textbf{The adoption of Artificial Intelligence (AI) in
  high-stakes domains such as healthcare, wildlife preservation,
  autonomous driving and criminal justice system calls for a
  data-centric approach to AI. Data scientists spend the majority of
  their time studying and wrangling the data, yet tools to aid them
  with data analysis are lacking. This study identifies the recurrent
  data quality issues in public datasets. Analogous to code smells, we
  introduce a novel catalogue of data smells that can be used to
  indicate early signs of problems or technical debt in machine
  learning systems. To understand the prevalence of data quality
  issues in datasets, we analyse 25 public datasets and identify 14
  data smells.}}

\end{abstract}

\maketitle

\section{Introduction}\label{sec:intro}
Data analysis is a critical and dominant stage of the machine learning
lifecycle. Once the data is collected, most of the work goes into
studying and wrangling the data to make it fit for training. A highly
experimental phase follows where a model is selected and tuned for
optimal performance. The final model is then productionised and
monitored constantly to detect data drifts and drop in
performance~\cite{sculley2015hidden,bosch2021engineering,hutchinson2021towards,sato2019continuous}.

When compared to traditional software, the feedback loop of a machine
learning system is longer. While traditional software primarily
experiences change in \textit{code}, a machine learning system matures
through changes in \textit{data, model \& code}
\cite{sato2019continuous}. Given the highly tangled nature of machine
learning systems, a change in any of the stages of the lifecycle
triggers a ripple effect throughout the entire pipeline
\cite{sculley2015hidden}. Testing such changes also becomes
challenging since all three components need to be tested. Besides the
traditional test suites, a full training-testing cycle is required
which incurs time, resource and financial costs. The surrounding
infrastructure of a machine learning pipeline becomes increasingly
complex as we move towards a productionised model. Thus catching
potential problems in the early, upstream phase of data analysis
becomes extremely valuable as fixes are faster, easier and cheaper to
implement.

AI has had a significant impact on the technology sector due to the
presence of large quantities of unbiased data~\cite{ng2021chat}. But
AI's true potential lies in its application in critical sectors such
as healthcare, wildlife preservation, autonomous driving, and criminal
justice system~\cite{crawford2021atlas}. Such high-risk domains almost
never have an existing dataset and require practitioners to collect
data. Once the data is collected, it is often small and highly
biased. While AI research is primarily dominated by model
advancements, this new breed of \textit{high-stakes AI} supports the
need for a more data-centric approach to
AI~\cite{kshirsagar2021becoming,sambasivan2021everyone,zhang2020machine}.

Since the study of software systems with machine learning components
is a fairly young discipline, resources are lacking to aid
practitioners in their day-to-day activities. The highly data-driven
nature of machine learning makes data equivalent to code in
traditional software. The notion of code smells is critical in
software engineering to identify early indications of potential bugs,
sources of technical debt and weak design choices. Code smells have
existed for over 30 years. A large body of scientific work has
catalogued the different smells, the context in which they occur and
their potential side-effects. To the best of our knowledge, such a
catalogue however does not exist for data science.

The research questions along with the contributions of this paper are
listed below.

\begin{itemize}
\item{\textbf{RQ1.}} \textbf{What are the recurrent data quality
  issues that appear in public datasets?}

  Analogous to code smells, we introduce the notion of data smells.
  Data smells are anti-patterns in datasets that indicate early signs
  of problems or technical debt.

\item {\textbf{RQ2.}} \textbf{What is the prevalence of such data quality issues in
  public datasets?}

  We create a catalogue of 14 data smells by analysing 25 popular
  public datasets\footnote{Our analysis of the datasets can be found
  on Figshare https://figshare.com/s/fd608796dd65f0808e7e}. The
  catalogue also presents real-world examples of the smells along with
  refactoring suggestions to circumvent the problem.

  Additionally, we plan to publish the catalogue online under the
  creative commons license in hopes that students and practitioners
  find it valuable.
\end{itemize}

The remainder of the paper is structured as follows. Section
\ref{sec:related} provides an overview of related concepts and prior
work that has been done. The methodology followed by this paper is
presented in Section \ref{sec:method} followed by the results in
Section \ref{sec:results}. The paper concludes with a discussion of
the results, limitations and future work in Section \ref{sec:discuss},
\ref{sec:limit} and \ref{sec:conclude} respectively.

\begin{figure*}
  \centering
  \includegraphics[width=0.75\linewidth]{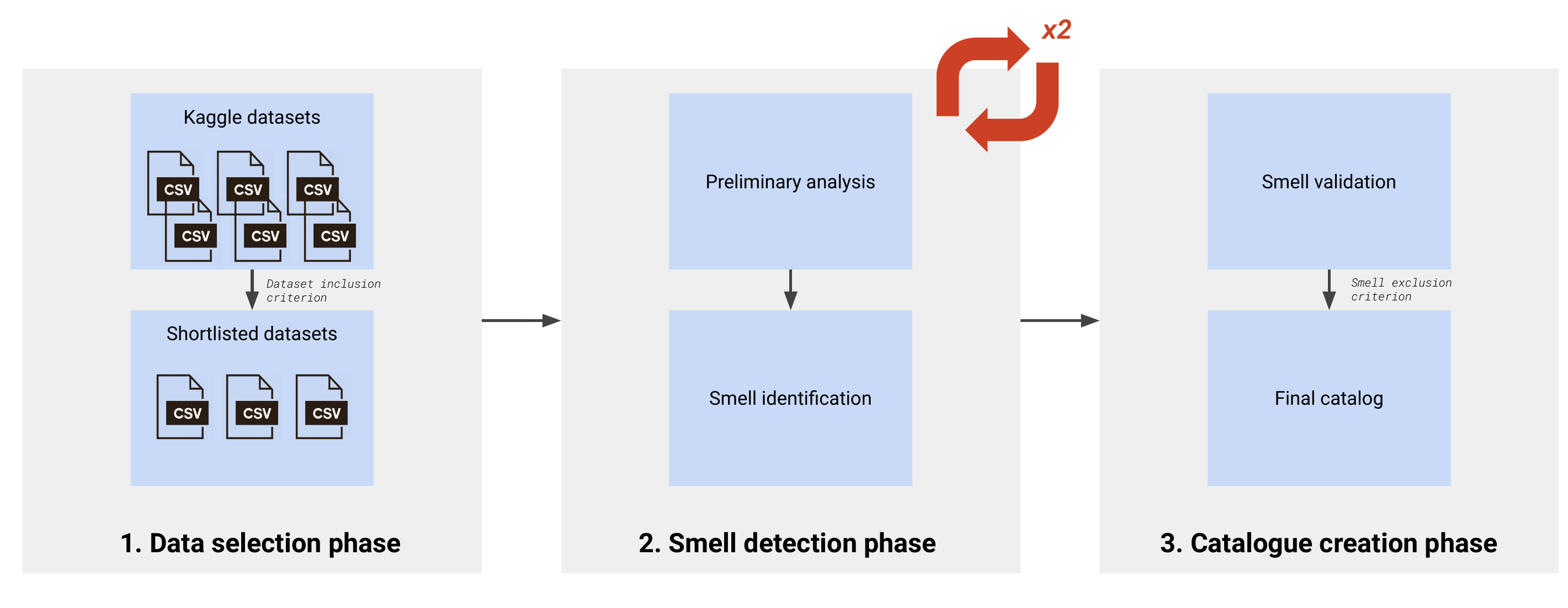}
  \caption{Overview of scientific process}
  \label{fig:method}
\end{figure*}

\section{Related Work}\label{sec:related}
This section provides an overview of relevant prior work in
\textit{code smells}, \textit{data validation} and \textit{AI
  engineering}.

Code smells were originally proposed by Kent Beck in the 1900s and
later popularised by \emph{\citeauthor{fowler2018refactoring}} in his
book \textbf{Refactoring}
\cite{fowler2018refactoring,fowler2006codesmell}. Code smells are
indications of potential problems in the code and require engineers to
investigate further. Common code smells include presence of bloated
code such as \textit{large classes} \& \textit{long methods},
redundant code such as \textit{duplicate code} \& \textit{dead code
  paths} and excessive coupling such as \textit{feature envy}
\cite{fowler2018refactoring,refactoringguru}. Code smells have been
widely adopted by the software engineering community to improve the
design and quality of their codebase. The notion of code smells has
also been extended to other areas such as
testing~\cite{bavota2015test,tufano2016empirical,spadini2018relation},
bug tracking~\cite{tuna2022bug}, code review~\cite{dougan2022towards}
and database management
systems~\cite{sharma2018smelly,muse2020prevalence,de2019prevalence}.
Code smells however still suffer from the problem of lacking
generalisability over a large population as most smells are subjective
to the developer, team or organisation.

Data validation is a well established field of research with roots in
Database Management Systems (DBMS). With the wide adoption of
data-driven decision-making by businesses, significant efforts have
been made towards automated data cleaning and quality
assurance~\cite{hellerstein2008quantitative,krishnan2016towards,chu2016data,schelter2018automating}. In
the context of machine learning, several tools and techniques have
been proposed for improving data quality and automated data
validation~\cite{saleem2014pre,krishnan2017boostclean,breck2019data,biessmann2021automated,zoller2021benchmark,lwakatare2021experiences}. \emph{\citeauthor{hynes2017data}}
present a data linting tool in the context of \emph{Deep Neural
Networks (DNNs)}. The tool checks the training data for potential
errors both at the dataset and feature level. The paper presents
empirical evidence of applying the linter to over 600 open source
datasets from Kaggle, along with several proprietary Google
datasets. The results indicate that such a tool is useful for new
machine learning practitioners and educational
purposes~\cite{hynes2017data}. Although there is some overlap between
the data linter by \citeauthor{hynes2017data} and our data smells
project, We argue that \citeauthor{hynes2017data} did not follow
a systematic approach to collect the linting rules. Our work is
complimentary to data linters as our approach exhaustively extracts
potential data quality issues from datasets. Our catalogue of data
smells can be seen as a framework for systematically extending, or
creating new data linting and validation tools.

AI engineering is a relatively young discipline of software
engineering (SE) research. The primary focus of the field is to
compare and contrast machine learning systems to traditional software
systems and adopt best practices from the SE community. The seminal
paper by \emph{\citeauthor{sculley2015hidden}} was the first to
recognise that machine learning systems accumulate technical debt
faster than traditional software~\cite{sculley2015hidden}. This
accelerated rate of technical debt accumulation is due to the highly
tangled nature of machine learning models to its data. Machine
learning is data-centric as each problem---which requires new or
combination of existing datasets---needs to be addressed
individually~\cite{sculley2015hidden,kim2017data,arpteg2018software,wan2019does,amershi2019software,bosch2021engineering}.

Data scientists spend the majority of their time working with data,
yet unlike in software engineering, lack tools that can aid them in
their analysis~\cite{bosch2021engineering,hutchinson2021towards}. This
study proposes a catalogue of data smells that can be beneficial to
practitioners and used as a framework for development of tools in the
future.

\begin{table*}
  \centering
  \caption{Selected Datasets}
  \begin{tabular}{p{0.1\linewidth} p{0.3\linewidth} r r r r c}
    \toprule
    \textbf{Name} & \textbf{Description} & \textbf{Size} &
    \textbf{Rows} & \textbf{Columns} & \textbf{Votes} &
    \textbf{Version} \\
    \midrule
    \textit{\href{https://www.kaggle.com/rodolfomendes/abalone-dataset}{abalone}}
    & Predicting age of abalone from physical measurements & 188K &
    4177 & 9 & 99 & 3 \\
    \textit{\href{https://www.kaggle.com/uciml/adult-census-income}{adult}}
    & Predicting whether income exceeds \$50K/yr based on census data
    & 3.8M & 32561 & 15 & 475 & 3 \\
    \textit{\href{https://www.kaggle.com/dgomonov/new-york-city-airbnb-open-data}{airbnb}}
    & Airbnb listings and metrics in NYC, NY, USA & 6.8M & 48895 & 16
    & 2502 & 3\\
    \textit{\href{https://www.kaggle.com/neuromusic/avocado-prices}{avocado}}
    & Historical data on avocado prices and sales volumn in Multiple
    US markets & 1.9M & 18249 & 13 & 2770 & 1 \\
    \textit{\href{https://www.kaggle.com/mczielinski/bitcoin-historical-data}{bitcoin}}
    & Bitcoin data at 1-min intervals from select exchanges, Jan 2012
    to March 2021 & 303M & 4857377 & 8 & 2876 & 7 \\
    \textit{\href{https://www.kaggle.com/uciml/breast-cancer-wisconsin-data}{breast-cancer}}
    & Predict whether the cancer is benign or malignant & 123K & 569
    & 33 & 2537 & 2 \\
    \textit{\href{https://www.kaggle.com/fivethirtyeight/fivethirtyeight-comic-characters-dataset}{comic-dc}}
    & FiveThirtyEight DC comic characters & 1.1M & 6896 & 13 & 2465 &
    111 \\
    \textit{\href{https://www.kaggle.com/fivethirtyeight/fivethirtyeight-comic-characters-dataset}{comic-marvel}}
    & FiveThirtyEight Marvel comic characters & 2.3M & 16376 & 13 &
    2465 & 111 \\
    \textit{\href{https://www.kaggle.com/gpreda/covid-world-vaccination-progress}{covid-vaccine}}
    & Daily and total vaccination for COVID-19 & 11M & 53595 & 15 &
    1978 & 234 \\
    \textit{\href{https://www.kaggle.com/gpreda/covid-world-vaccination-progress}{covid-vaccine-manufacturer}}
    & Vaccinations for COVID-19 by manufacturer & 793K & 19168 & 4 &
    1978 & 234 \\
    \textit{\href{https://www.kaggle.com/usgs/earthquake-database}{earthquake}}
    & Date, time and location of all earthquakes with magnitude of 5.5
    or higher & 2.3M & 23412 & 21 & 435 & 1 \\
    \textit{\href{https://www.kaggle.com/mlg-ulb/creditcardfraud}{fraud}}
    &
    Anonymised credit card transactions labeled as fraudulent or
    genuine & 144M & 284807 & 31 & 8775 & 3 \\
    \textit{\href{https://www.kaggle.com/unsdsn/world-happiness}{happiness}}
    & Happiness scored according to economic production, social
    support, etc & 8.7K & 156 & 9 & 3401 & 2 \\
    \textit{\href{https://www.kaggle.com/ronitf/heart-disease-uci}{heart}}
    & UCI heart disease dataset & 12K & 302 & 14 & 5601 & 1 \\
    \textit{\href{https://www.kaggle.com/mirichoi0218/insurance}{insurance}}
    & Insurance forecast by using linear regression & 55K & 1338 & 7
    & 1621 & 1 \\
    \textit{\href{https://www.kaggle.com/uciml/iris}{iris}} & Classify
    iris plants into three species & 4.5K & 150 & 5 & 2779 & 2 \\
    \textit{\href{https://www.kaggle.com/shivamb/netflix-shows}{netflix}}
    & Listings of movies and tv shows on Netflix & 3.3M & 8807 & 12 &
    6155 & 5 \\
    \textit{\href{https://www.kaggle.com/aparnashastry/building-permit-applications-data}{permit}}
    & San Francisco building permits & 76M & 198900 & 37 & 194 & 1 \\
    \textit{\href{https://www.kaggle.com/mdp1990/google-play-app-store-eda-data-visualisation/data}{playstore}}
    & Google play store apps data & 1.3M & 10841 & 13 & 47 & 1 \\
    \textit{\href{https://www.kaggle.com/spscientist/students-performance-in-exams}{student}}
    & Marks secured by students in various subjects & 71K & 1000 & 8
    & 3050 & 1 \\
    \textit{\href{https://www.kaggle.com/russellyates88/suicide-rates-overview-1985-to-2016}{suicide}}
    & Suicide rates overview 1985 to 2016 & 2.6M & 27820 & 12 & 2766 &
    1 \\
    \textit{\href{https://www.kaggle.com/blastchar/telco-customer-churn}{telco}}
    & Telco customer churn & 955K & 7043 & 21 & 1902 & 1 \\
    \textit{\href{https://www.kaggle.com/gregorut/videogamesales}{vgsales}}
    & Video game sales & 1.3M & 16598 & 11 & 4248 & 2\\
    \textit{\href{https://www.kaggle.com/uciml/red-wine-quality-cortez-et-al-2009}{wine}}
    & Red wine quality & 11K & 178 & 14 & 1918 & 2\\
    \textit{\href{https://www.kaggle.com/datasnaek/youtube-new}{youtube}}
    & Trending YouTube video statistics & 60M & 40949 & 16 & 4381 & 115\\
    \bottomrule
  \end{tabular}
  \label{tab:datasets}
\end{table*}

\section{Methodology}\label{sec:method}
Figure~\ref{fig:method} presents an overview of the scientific process
followed in this study. The methodology can be divided into three
distinct phases which are presented in more detail below.

\subsection{Data selection phase}~\label{sec:method-select}
This study uses Kaggle---an online data repository---to discover
datasets for the analysis~\footnote{https://www.kaggle.com}. All
public datasets available on Kaggle are sorted by the \emph{Most
Votes} criteria and 25 datasets are shortlisted based on the inclusion
criterion presented in Table~\ref{tab:dataset-ic}. The sample of
datasets only includes CSV files of size smaller than 1GB to
facilitate the analysis on a personal laptop. Analysis of unstructured
datasets such as text corpus, images, videos and audio is excluded as
this calls for specialised tools, additional time and effort which was
deemed beyond the scope of this study. Structured datasets on the
other hand are more commonly occurring. Practitioners and academics
frequently work with structured datasets that are often used for
educational purposes. Therefore, we invested our efforts in analysing
structured datasets to make our work relevant to a larger demographic.
The 25 datasets selected for this study are listed in
Table~\ref{tab:datasets}. The table also includes additional metadata
such as the \emph{size, number of rows and columns, number of votes}
and the \emph{latest version} at the time of analysis.

\begin{table}
  \centering
  \caption{Inclusion criterion for datasets}
  \begin{tabular}{p{0.1\linewidth} p{0.7\linewidth}}
    \toprule
    \textbf{Key} & \textbf{Inclusion Criteria} \\
    \midrule
    \emph{IC1} & Dataset is of \emph{CSV} format. \\
    \emph{IC2} & Dataset is smaller than 1GB in size. \\
    \emph{IC3} & Dataset contains structured data. \\
    \emph{IC4} & Dataset primarily contains numerical and categorical
    features. \\
    \bottomrule
  \end{tabular}
  \label{tab:dataset-ic}
\end{table}

\subsection{Smell detection phase}~\label{sec:method-detect}

We used the Python programming
language~\footnote{https://www.python.org} along with the data
analysis package Pandas~\footnote{https://pandas.pydata.org} to
perform the analysis. The smells are identified using a two pass
technique which is manually conducted by the first author. The first
pass focuses on identifying characteristics of datasets that are
indicative of a smell. Since the catalogue of smells evolved as more
datasets were analysed during the first pass, a second pass is used to
validate the original smells observed in the datasets. The second pass
also helps to identify newer smells which were missed in the older
datasets during the first pass.

The first pass conducts a preliminary analysis of the datasets listed
below. This is a standard list of checks performed by data scientists
during data understanding under the CRISP-DM model of data
mining~\cite{ibm-crisp,schroer2021systematic,saltz2021crisp,martinez2019crisp}.

\begin{enumerate}
  \item Reading the accompanying data documentation when available.
  \item Analysing their \emph{head} and \emph{tail}---both in its
    entirety and on a feature-by-feature basis.
  \item Observing the column headers and datatypes for relevant meta
    data such as the expected schema of the dataset.
  \item Analysing the descriptive statistics of the dataset.
  \item Checking for missing values and duplicate rows.
  \item And finally checking correlations amongst features.
\end{enumerate}

We did not analyse the distribution of the features and their
relationship with one another (besides checking for correlation). This
is because the insights gained from distributional and relational
analysis of a particular dataset are not generalisable to other
datasets and domains. This is touched upon in more detail in
Section~\ref{sec:limit}.

\subsection{Catalogue creation phase}
We further prune the list of smells using the exclusion criterion
listed in Table~\ref{tab:smell-ec} and additional validation from the
second author. Smells which cannot be generalised to other structured
datasets are removed. Similarly, smells which are relevant only when
using a specific programming language and tools (such as Python and
Pandas) but not applicable when using a different toolset (such as
Matlab~\footnote{https://www.mathworks.com/products/matlab.html},
R~\footnote{https://www.r-project.org/} or
Julia~\footnote{https://julialang.org/}) are excluded.

\begin{table}
  \centering
  \caption{Exclusion criterion for smells}
  \begin{tabular}{p{0.1\linewidth} p{0.7\linewidth}}
    \toprule
    \textbf{Key} & \textbf{Exclusion Criteria} \\
    \midrule
    \emph{EC1} & Smell is not generalisable to structured datasets. \\
    \emph{EC2} & Smell is not generalisable to other programming
    languages and tools. \\
    \bottomrule
  \end{tabular}
  \label{tab:smell-ec}
\end{table}

\section{Results}\label{sec:results}
This section presents the results obtained from the analysis of public
datasets. The most recurrent data quality issues are presented first.
A catalogue of data smells showing the prevalence of such data quality
issues is presented next (See RQ1 and RQ2 in Section~\ref{sec:intro}).
This study analysed \textbf{25} public datasets from which \textbf{14}
data smells were discovered. We group the smells into 4 distinct
categories based on their similarity as listed below.

\begin{enumerate}
\item \textbf{Redundant value smells} or smells which occur due to
  presence of features that do not contribute any new information.
\item \textbf{Categorical value smells} or smells which occur due to
  presence of features containing categorical data.
\item \textbf{Missing value smells} or smells which occur due to
  absence of values in a dataset.
\item \textbf{String value smells} or smells which occur due to
  presence of features containing string type data.
\end{enumerate}

Additionally, three more smells were found which could not be grouped
into the above categories and are put under the \textbf{Miscellaneous
  smells} category.

\begin{table}[htb]
  \centering
  \caption{List of smells}
  \resizebox{\linewidth}{!}{%
  \begin{tabular}{l l r l}
    \toprule
    \textbf{Key} & \textbf{Name} & \multicolumn{2}{l}{\textbf{Count}}\\
    \midrule
    \multicolumn{2}{l}{\emph{\textbf{Redundant value smells (red)}}} & 33 & \catbar{33}\\
    \midrule
    \emph{red-corr} & Correlated features & 19 & \subcatbar{19}\\
    \emph{red-uid} & Unique identifiers & 11 & \subcatbar{11} \\
    \emph{red-dup} & Duplicate examples & 3 & \subcatbar{3} \\
    \midrule
    \multicolumn{2}{l}{\emph{\textbf{Categorical value smells (cat)}}} & 17 & \catbar{17}\\
    \midrule
    \emph{cat-hierarchy} & Hierarchy from label encoding & 12 & \subcatbar{12}\\
    \emph{cat-bin} & Binning categorical features & 5 & \subcatbar{5}\\
    \midrule
    \multicolumn{2}{l}{\emph{\textbf{Miscellaneous value smells (misc)}}} & 14 & \catbar{14}\\
    \midrule
    \emph{misc-unit} & Unknown unit of measure & 9 & \subcatbar{9}\\
    \emph{misc-balance} & Imbalanced examples  & 3 & \subcatbar{3}\\
    \emph{misc-sensitive} & Presence of sensitive features & 2 & \subcatbar{2}\\
    \midrule
    \multicolumn{2}{l}{\emph{\textbf{Missing value smells (miss)}}} & 13 & \catbar{13} \\
    \midrule
    \emph{miss-null} & Missing values & 11 & \subcatbar{11}\\
    \emph{miss-sp-val} & Special missing values & 1 & \subcatbar{1} \\
    \emph{miss-bin} & Binary missing values & 1 & \subcatbar{1}\\
    \midrule
    \multicolumn{2}{l}{\emph{\textbf{String value smells (str)}}} & 12 & \catbar{12}\\
    \midrule
    \emph{str-num} & Numerical feature as string & 5 & \subcatbar{5} \\
    \emph{str-sanitise} & Strings with special characters & 5 & \subcatbar{5}\\
    \emph{str-human} & Strings in human-friendly formats & 2 & \subcatbar{2}\\
    \bottomrule
  \end{tabular}}
  \label{tab:smells}
\end{table}

Table~\ref{tab:smells} presents an overview of all data smells along
with their distribution. The remainder of this report frequently
refers to these smells by their unique key which is also provided
here. The \emph{redundant and categorical value smells} are the most
common categories with a total occurrence of 33 and 17 respectively.
The \emph{missing \& string value smells} are the least common
categories with a total occurrence of 13 and 12 respectively.
\emph{red-corr} is the most common smell, observed in 19 of the 25
datasets. The remaining top five smells include \emph{cat-hierarchy,
miss-null, red-uid and misc-unit} which are observed in more than 10
datasets. The least frequently observed smells include
\emph{misc-balance, str-human, misc-sensitive, miss-sp-val and
miss-bin} which are observed in less than 5 datasets.

\begin{figure}
  \centering
  \includegraphics[width=0.725\linewidth]{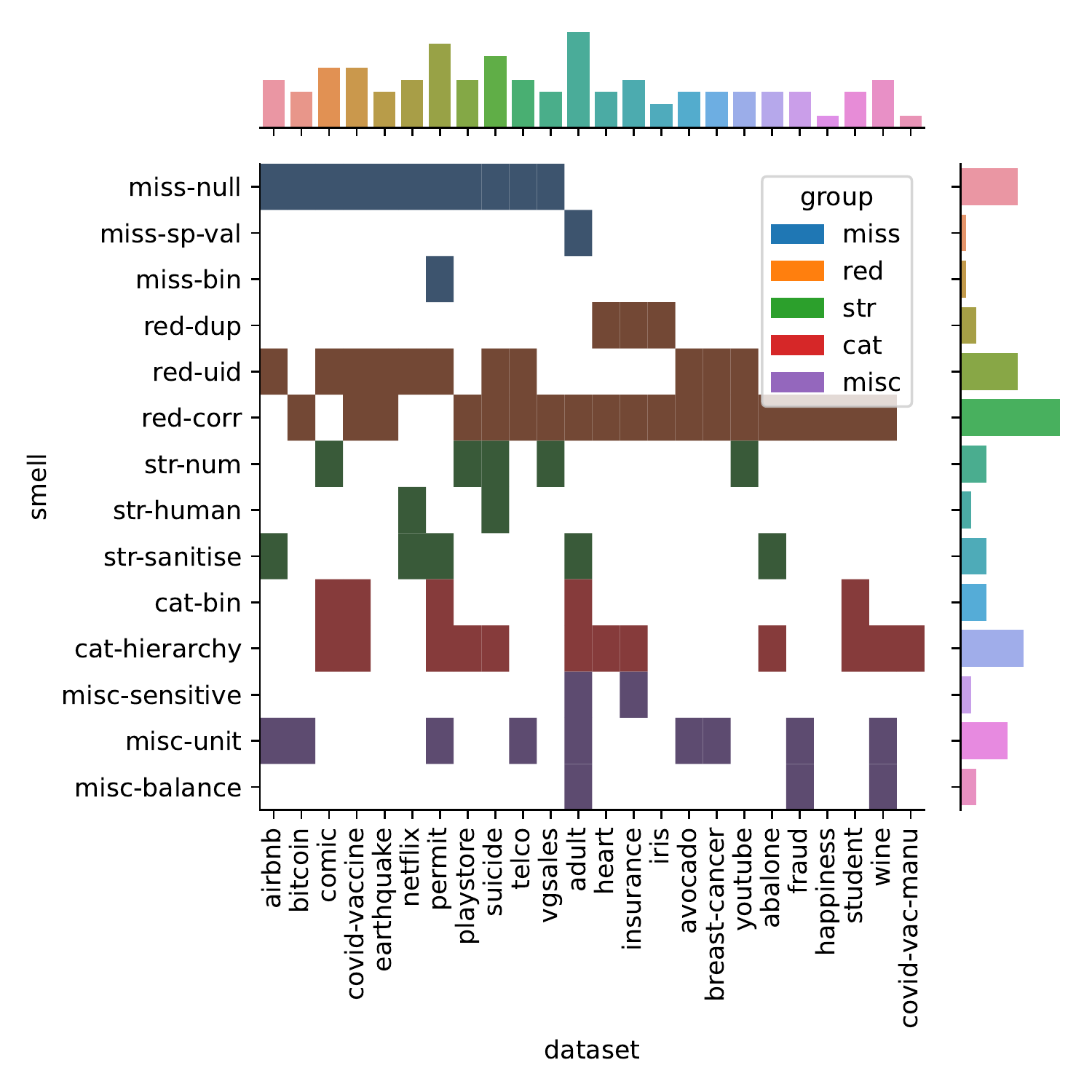}
  \caption{Joint distribution of smells and datasets}
  \label{fig:jointplot@meta--dataset-smell-hue:group}
\end{figure}

Finally we also analyse the distribution of smells within the
datasets. Figure~\ref{fig:jointplot@meta--dataset-smell-hue:group}
shows a two dimensional histogram of the smells and datasets such that
the intersection of datasets where a particular smell occurred is
filled. The histogram is colour-coded based on the smell category
which allows us to observe the most common smell categories at a
glance. The figure also contains two marginal plots across the x and y
axes. The marginal plot along the x axis presents a count of smells
\emph{\textbf{within}} each dataset such that we can identify the
datasets with the most and least number of smells. Similarly, the
marginal plot along the y axis presents a count of smells
\emph{\textbf{across}} all datasets such that we can identify the most
and least frequently occurring smells.

The remainder of this section presents the smell groups and their
corresponding smells in more detail. We present examples of the smells
discovered along with an explanation of the underlying problems that
may arise. Where applicable, potential strategies to mitigate the
problem, context in which the smells may not apply and references from
literature (both scientific and grey) are also presented.

\subsection{Redundant Value Smells}\label{sec:results-red}
This section presents smells that indicate presence of redundant
information in datasets. This was the most prominent group of smells
as 33 occurrences were observed in this study.

\subsubsection{Correlated features (red-corr)}\label{sec:results-red-corr}
This study identified 19 datasets that contained correlated features.

Checking for correlation amongst features is a common practise in data
science. The correlation between two numerical features is
a representation of the linear relationship between them. From
a machine learning perspective, presence of uncorrelated features
indicates that the features impart new information. But the presence
of correlated features is a smell for redundant information.

Datasets come in all shapes and sizes, often containing rows or
columns which do not offer new or valuable information. The model
training stage usually helps identify features which do not need to be
included in the training data, allowing us to engineer a more
efficient dataset. Engineering efficient datasets is important since
the machine learning lifecycle consists of several deeply coupled
stages. Naturally, any form of optimisation---no matter how
small---propagates through the downstream stages. A small dataset is
easier to understand, faster to train a model on and takes up less
storage. The benefits of a small dataset are appreciated especially
during the model development phase where many experiments along with
their accompanying dataset, model and code are versioned and
stored~\cite{sato2019continuous,arpteg2018software}.

Presence of correlated features gives practitioners the opportunity to
perform feature selection and drop redundant features which do not
affect the model's performance.

\subsubsection{Unique identifiers (red-uid)}\label{sec:results-red-uid}
This study identified 11 datasets that contained columns containing a
unique identifier (uid) for each example in the dataset. For example,
the \emph{youtube, earthquake, netflix, telco and avocado} datasets
all contain a column carrying a uid for the examples.

Relational databases are the most popular type of databases being used
today. Such databases have a column containing a unique identifier
(uid) commonly referred to as the \emph{primary key}. Machine learning
pipelines are often automated and perform end-to-end operations,
starting with data consumption from large data warehouses and data
lakes, all the way to publishing a trained machine learning model in
production. Although uids are useful when performing merge or join
operations on two or more database tables, they become redundant when
training machine learning models. Their presence in a dataset is a
smell for potential problems in downstream stages.

A machine learning model may learn some hidden relationship between
the uids and the target values that produces a high accuracy during
training. Such an insight however prevents the model from learning
relationships and trends that are generalisable to unseen data and
limits its ability to provide meaningful predictions. Furthermore,
uids may also prevent the detection of duplicate examples in a
dataset. This is another smell that was discovered in this study and
discussed further in Section~\ref{sec:results-red-dup} below.

Although features containing uids in general should not be included in
the training set, they can sometimes provide valuable insights during
the data analysis stage. The \textit{airbnb} dataset contains the
\texttt{host\_id} and \texttt{id} features containing uids for the
hosts and the properties respectively. One may regard them as
redundant feature and drop them. However further analysis of the
columns may lead to interesting insights. For instance, the
\texttt{host\_id} column contains duplicate entries which presents the
insight that hosts may own multiple properties. This can further be
engineered into a new feature which may help during training.
Analysing the columns together can help detect truly duplicate
examples (rows with the same property and host id) and outliers (rows
with the same property id but different host ids).

\subsubsection{Duplicate examples (red-dup)}\label{sec:results-red-dup}
This study identified three datasets, namely \emph{heart, insurance
and iris}, that contained duplicate rows. We ignore timeseries data
where an event can occur several times resulting in duplicate rows.

Duplicate examples in a dataset are defined as two or more rows which
refer to the same entity. They do not serve any purpose and can be
removed from the dataset, making their presence a smell for redundancy
in the dataset.

Duplicate examples make a dataset bloated. They do not contribute any
new information during the data analysis stage. Furthermore, training
a machine learning model with a dataset containing duplicate examples
can impede the model's performance on unseen data. Training a model
using duplicate examples might lead to \textit{overfitting} as it may
learn once from the original example and then again from the duplicate
example(s).

\subsection{Categorical Value Smells}\label{sec:results-cat}
This section presents smells that arise from the presence of categorical
data. This study found 17 occurrences of smells from this group.

\subsubsection{Hierarchy from label encoding (cat-hierarchy)}\label{sec:results-cat-hierarchy}
\begin{figure}
  \centering
  \includegraphics[width=0.725\linewidth]{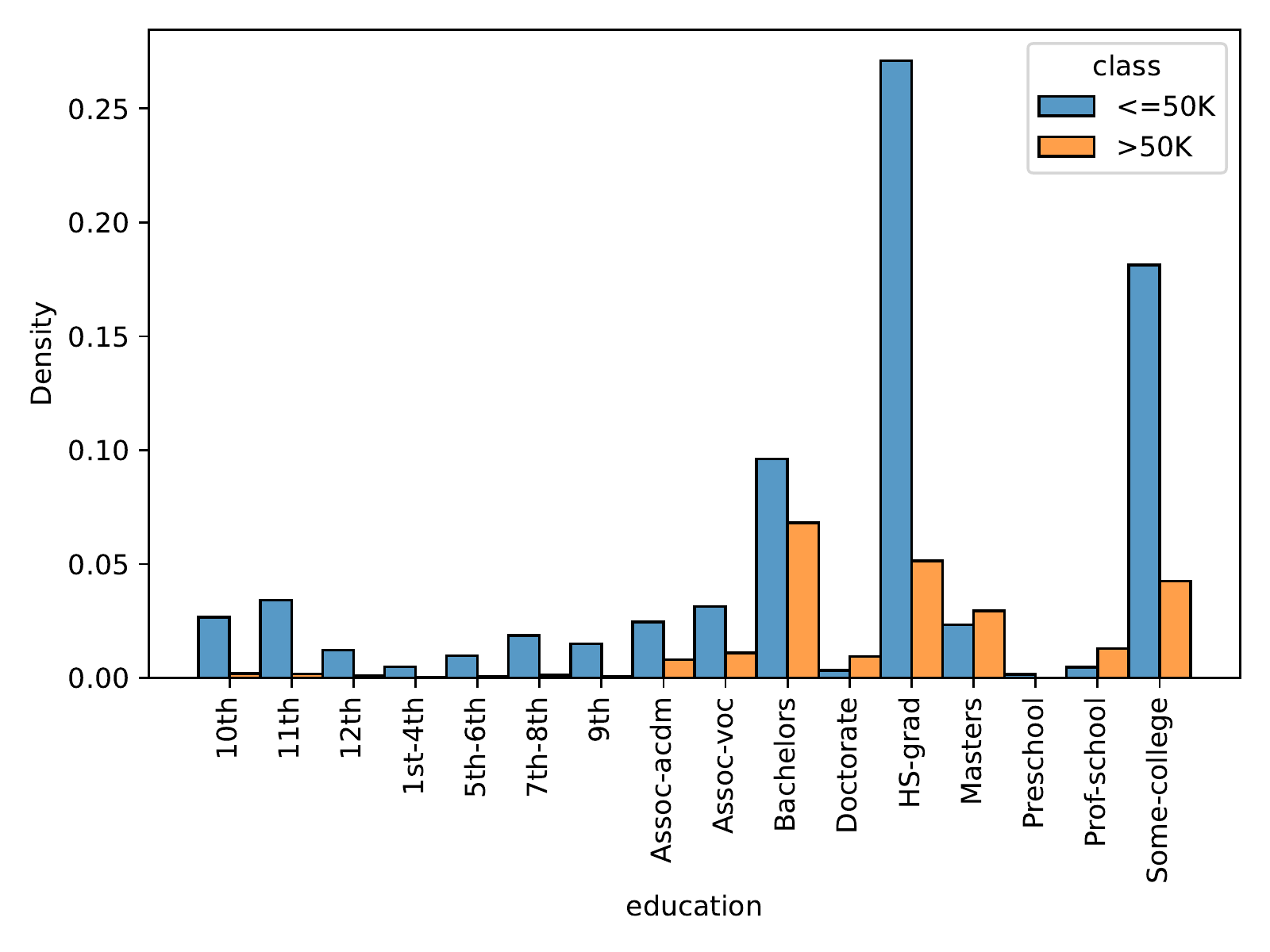}
  \caption{Probability density plot of adult education based on income}
  \label{fig:histplot@adult--education-hue:class}
\end{figure}

The values of the \texttt{education} feature in the \emph{adult}
dataset have a clear hierarchy amongst themselves.
Figure~\ref{fig:histplot@adult--education-hue:class} shows the
probability density plot of the education levels of adults conditioned
on their income class. For the given dataset, the probability that an
adult earns more given that they have better education is higher. We
can expose the hierarchy amongst the education levels by assigning a
number starting from 0 in ascending order such that higher levels of
education are assigned a higher numerical value. This encoding scheme
can aid a machine learning model to accurately predict the income
class of an individual. Using the same encoding scheme for the
\texttt{sex} and \texttt{race} features in the same dataset can
however lead to biased outcomes.

An important characteristic of categorical data is the notion of
hierarchy amongst its values. \emph{Label or dummy encoding} is a
common technique used in data science to encode categorical data as
numbers. This technique preserves the hierarchy amongst the values
which may impart useful information to the model during training. Some
categorical features however contain sensitive information (sensitive
features are discussed in more detail in
Section~\ref{sec:results-misc-sensitive}) and do not have hierarchy
amongst their values. Label encoding such features can introduce bias
into a machine learning model and affect its performance. The presence
of sensitive categorical features are therefore a smell to avoid
introducing bias into the model.

Label encoding sensitive categorical features can introduce unwanted
hierarchy amongst the values and lead to incorrect and biased results
in machine learning models. The model may incorrectly associate a sex
or race with a higher numerical value to be superior to other values
with a lower number. This can be avoided using the \emph{one-hot
encoding} technique as opposed to label encoding~\cite{al2021insider}.

\subsubsection{Binning categorical features (cat-bin)}\label{sec:results-cat-bin}
\begin{figure*}
  \centering
  \includegraphics[width=0.725\linewidth]{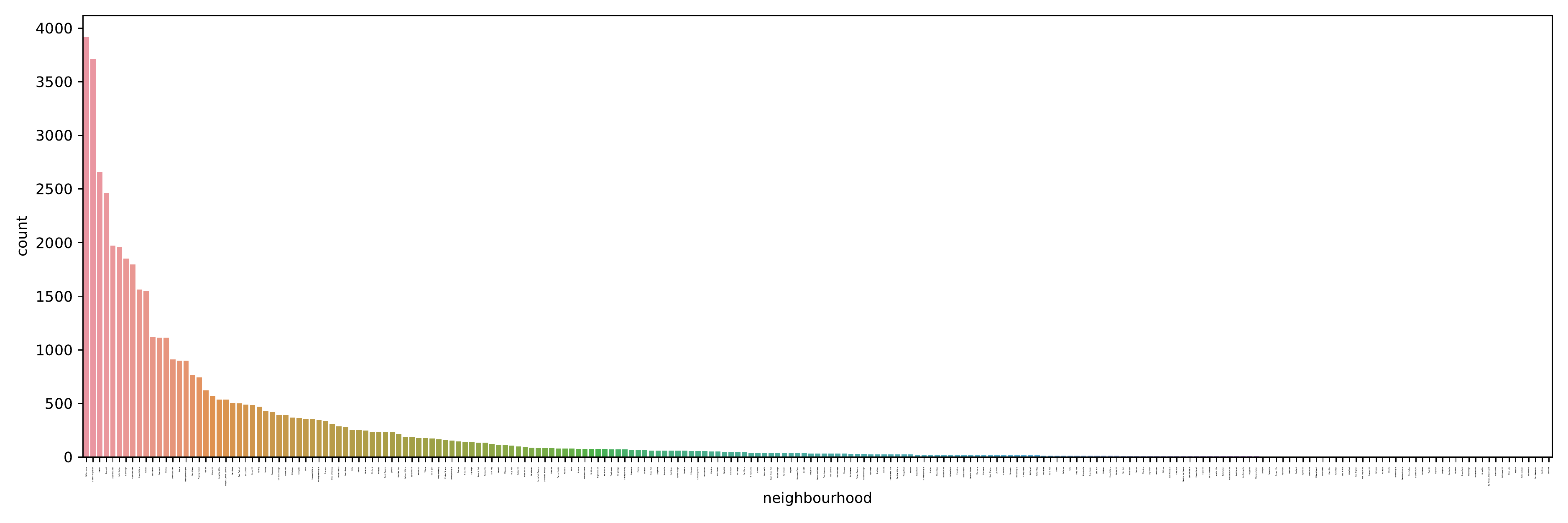}
  \caption{Histogram of \texttt{neighbourhoods} in \emph{airbnb}
    dataset with high cardinality.}
  \label{fig:airbnb-bin}
\end{figure*}

Figure~\ref{fig:airbnb-bin} presents the distribution of the
\texttt{neighbourhood} feature in the \emph{airbnb} dataset. It is a
categorical feature that contains over 200 unique values but several
values are rare and do not occur that often. Another example is found
in the \emph{adult} dataset where the \texttt{native-country} feature
contains 42 unique values.

One-hot encoding a feature with high cardinality can result in a very
large feature space and incur higher memory, disk space and
computation costs throughout the machine learning lifecycle. Presence
of categorical features with high cardinality in their data is a smell
to perform potential data transformations to reduce the cardinality.

\sloppypar{A common practise amongst data scientists to address such a
  problem is to bin several values together. For example, the
  \texttt{native-county} values can be binned into the seven
  continents. As an alternative to the \texttt{neighbourhood} feature,
  the \emph{airbnb} dataset also contains the
  \texttt{neighbourhood\_group} feature which bins the neighbourhood
  values into 5 broader areas.}

\subsection{Miscellaneous Smells}\label{sec:results-misc}
This section presents three smells which did not fit into groups
presented earlier. This study identified 14 occurances of smells from
this group.

\subsubsection{Presence of sensitive features (misc-sensitive)}\label{sec:results-misc-sensitive}
\begin{figure*}
  \centering
  \subcaptionbox{Probability density plot of adult income based on
    their race (along x axis) and sex (along y
    axis).\label{fig:histplot@adult--sex-race}}{\includegraphics[width=0.425\linewidth]{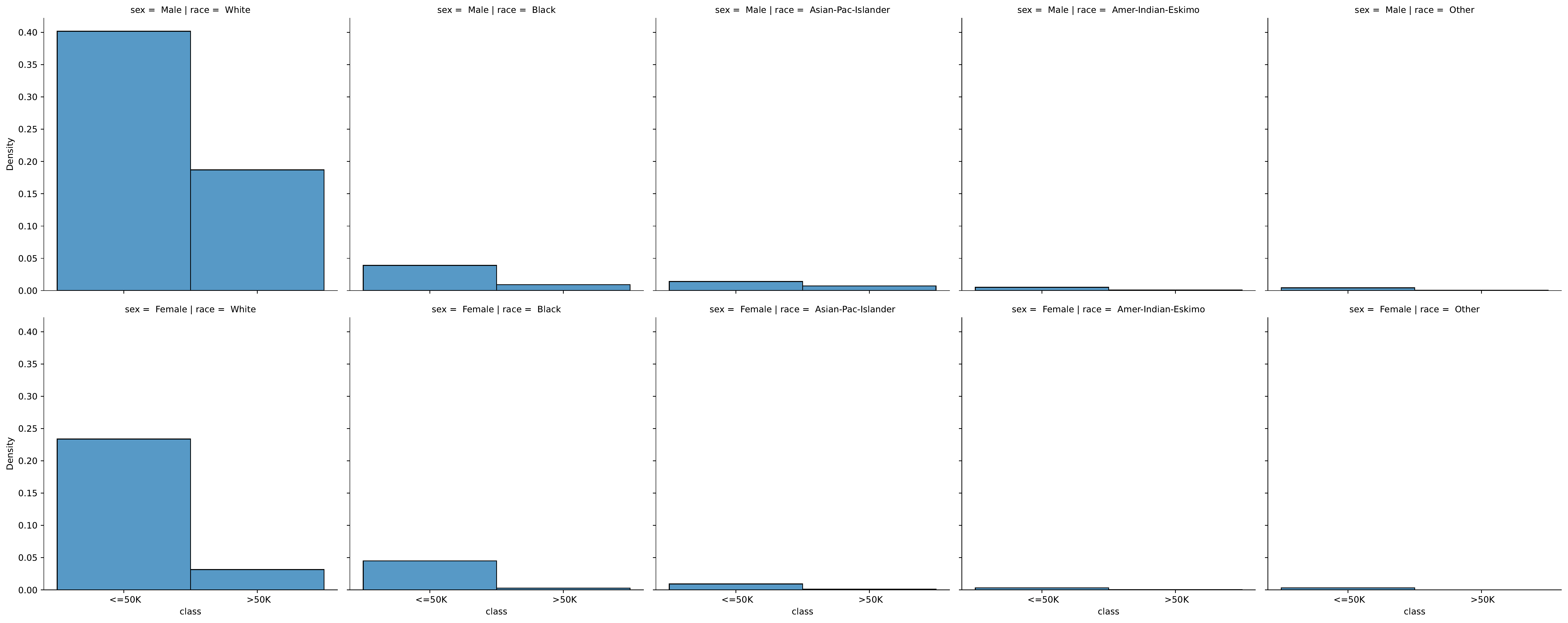}}
  \subcaptionbox{Histogram of \texttt{class} feature from \emph{fraud}
    dataset.\label{fig:fraud-class}}{\includegraphics[width=0.225\linewidth]{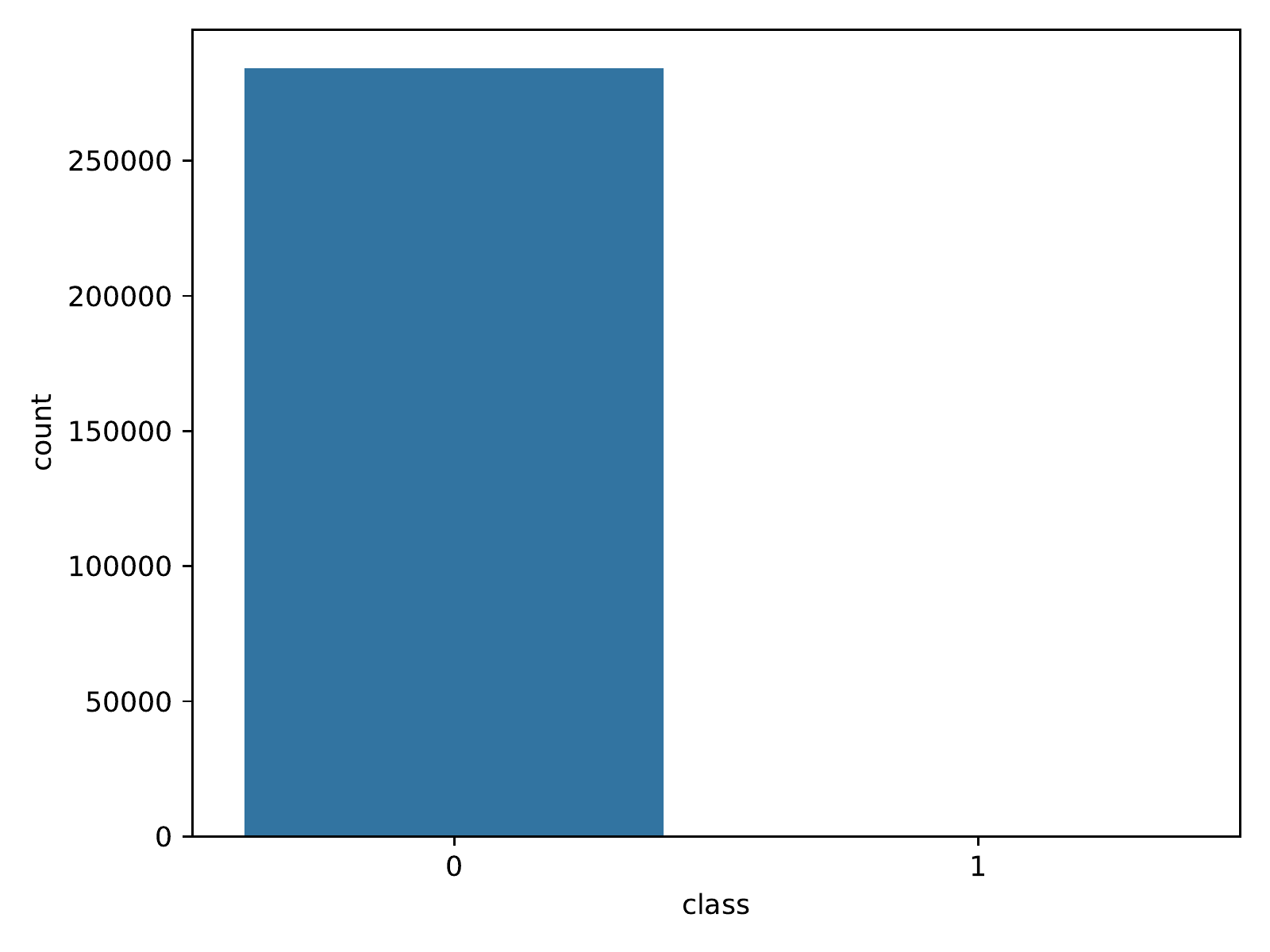}}
  \caption{Distribution of sensitive features from the \emph{adult}
    and \emph{fraud} datasets.}
\end{figure*}

The \textit{adult} dataset presents census information of individuals
from 1994. Amongst others, the dataset contains information regarding
the \textit{sex, race \& income} of individuals.
Figure~\ref{fig:histplot@adult--sex-race} presents the probability
density plot of the income class conditioned on the race and sex of
individuals. We see that for this dataset, the probability that a male
of lighter skin earns more than their female and darker skin
counterpart, is significantly higher.

Not all features contribute equally towards knowledge, whether it be
during the analysis or model training. Section~\ref{sec:results-red}
motivated the need to remove redundant features from a dataset,
leaving behind \emph{high-impact} features that contribute the most
towards analysis and model training. While most high-impact features
lead to interesting insights during analysis, not all should be used
to train machine learning models. Presence of high-impact features are
a smell to identify sensitive features that may lead to biased and
unfair model predictions.

Going back to the example presented above, a machine learning model
trained and tested on this dataset would be able to predict the income
class of an individual with high accuracy. However such a model when
used in production to making business decisions will lead to unfair
and biased predictions given it was trained with historical data with
similar
traits~\cite{miller2020is,weber2020black,berkeley2018mortgage}. Use of
biased models for predictive policing and criminal justice systems can
have far more devastating
consequences~\cite{heaven2020predictive,lifshitz2021racism,mesa2021can}.

Potential mitigation strategies include not using sensitive features
during model training and introducing appropriate regularisation
techniques to combat the bias. For instance, in the \emph{adult}
dataset, the \texttt{sex} and \texttt{race} can be excluded from the
training set so that the model can learn from more generalisable
features such as the age and the level of education of an individual.
More recently, the trustworthy AI research field has also seen
significant developments to address issues regarding safety and
robustness, explainability, fairness and privacy in machine learning
models~\cite{liu2021trustworthy}.

\subsubsection{Imbalanced examples (misc-balance)}\label{sec:results-misc-balance}
The \textit{fraud} dataset contains anonymised information on credit
card transactions for over $200,000$ European cardholders. The dataset
also contains a binary \texttt{class} feature where fraudulent
transactions are assigned a value of 1 and others a value of 0. As
seen from Figure~\ref{fig:fraud-class}, a model trained on this
dataset to detect fraudulent transactions will not perform well as the
dataset contains very few examples for the fraudulent transactions
class.

When performing classification using supervised learning algorithms,
the target feature can also contribute to bias in the model. A special
case of the \emph{misc-sensitive} smell is the presence of unbalanced
examples for the classes in a dataset.

Common techniques for working with skewed datasets include using more
robust performance metrics such as precision and recall, increasing
the size of the training set or manipulating the training set such
that the number of examples per class is equal.

\subsubsection{Unknown unit of measure (misc-unit)}\label{sec:results-misc-unit}
The \textit{breast-cancer} dataset contains several numerical features
such as \texttt{radius, perimeter \& area} of tumours. However the
column names and documentation fail to mention the unit in which the
features were measured.

Numerical data or data of the type \texttt{int} or \texttt{float} are
the most common data types in structured datasets. Understanding the
distribution and trends within numerical features constitutes the bulk
of the data analysis stage. This helps in gaining a deep understanding
of its characteristics which is vital to determine how a model will
perform when trained on such a feature set. Consistency is of high
importance when working with numerical features, and lack of a common
unit of measure for all observations of a feature is an early
indicator of problems during model training.

Lack of standardised data collection procedure, long durations of data
collection and use of undocumented data sources can all lead to
observations measured in different units. A mix of units can also lead
to incorrect results from outlier detection and propagate to
engineered features. Mean removal and variance scaling are common
pre-processing techniques used to standardise the numerical features
of a dataset prior to training a model. But such a transformation is
unfruitful if the observations are not recorded with the same unit of
measure.
\subsection{Missing Value Smells}\label{sec:results-miss}
This section presents smells that arise from presence of missing
values in datasets. This study identified 13 occurrences of smells
from this group.

\subsubsection{Nulltype missing values (miss-null)}\label{sec:results-miss-null}
A common problem observed in the analysed datasets is the absence of
data which can be an early indicator of potential problems in
downstream processes. We noticed certain abnormalities while analysing
the descriptive statistics of the \emph{permit, bitcoin and
covid-vaccine} datasets. Subsequently, the missing data check revealed
that these datasets had large quantities of missing values. For
instance, the \textit{permit} dataset contains several datetime
features where 50\% of the data is missing; 25\% of the data in the
\textit{bitcoin} dataset; and 50\% of the data in the numerical
features of the \textit{covid-vaccine} dataset are missing.

When a small portion of the data is missing or if sufficient number of
examples for each class is available, missing data can be omitted
prior to further analysis~\cite{aljuaid2016proper}. Such a strategy
ceases to be an option in high-stakes domains where data is limited
and highly unbalanced to begin with. In such datasets, dropping
missing data amplifies the imbalance and leads to insufficient data
for training machine learning
models~\cite{sambasivan2021everyone,kshirsagar2021becoming}. Missing
values are typically ignored by data analysis tools while performing
statistical computations (such as descriptive statistics), leading to
inaccurate and biased conclusions. Missing values can also lower the
performance of machine learning models due to underrepresented groups
in the dataset~\cite{nugroho2019missing,marlin2008missing}. The
problem gets worse as we scale to larger datasets which do not fit in
the memory of a single computer and require a more distributed
approach for storage and performing transformations.

Such cases require further effort from data scientists to impute the
missing values. Imputation of missing data is a vast research field in
itself with techniques ranging from simple statistical techniques such
as mean, median and linear regression, to using machine learning
models that predict the missing
value~\cite{aljuaid2016proper,saleem2014pre,twala2005comparison}.

\subsubsection{Special missing values (miss-sp-val)}\label{sec:results-miss-sp-val}
The missing values in the \emph{adult} dataset are represented with
the question mark character. Although the null type (\texttt{null} or
\texttt{nil}) is the most common data type used to represent missing
values, sometimes special string characters and keywords such as
\texttt{`?', `nil', `null'} are also used. \emph{Dummy encoding} is
another popular technique where a unique numerical value (such as
\texttt{-9999} or \texttt{-6666}) which is unlikely to be observed in
real-life is used. Using special characters, keywords and numbers to
represent missing values---especially when they are undocumented---are
a smell for problems in downstream stages.

Data analysis tools often contain built-in functionality to check for
presence of missing values by detecting null data types in the
dataset. Unless otherwise documented, using special characters or
numbers impedes the ability of data scientists and data analysis tools
to detect missing values accurately. We can indirectly discover string
type missing values when performing statistical computations. In such
a case, the data analysis tool will fail to perform a numerical
operation on data represented as a string and raise an error. Usage of
undocumented dummy encoding is however worse as the data analysis tool
will continue to operate. This adds to technical debt and may result
in incorrect statistical conclusions or catastrophic failure in
downstream stages. The data must be manually analysed to identify the
special character or number used to represent missing values, wasting
time and effort that could have been used to perform other productive
tasks.

Using null data types to represent missing values during data
collection phase is regarded as good practise. Using special
characters, keywords or dummy encoding for missing values must be
documented to reduce technical debt and aid future practitioners.

\subsubsection{Binary missing values (miss-bin)}\label{sec:results-miss-bin}
\sloppypar{While analysing the \emph{permit} dataset, we identified
  two features (namely \texttt{structural\_notification} and
  \texttt{tidf\_compliance}) with 90\% missing values. All non-missing
  values however comprised of the string \texttt{`Y'}, a common
  abbreviation for \emph{`yes'}. This indicated that the missing
  values carried an implicit meaning of \texttt{`N'} or \emph{`no'}
  and were not indicative of truly missing values.}

In Section~\ref{sec:results-miss-null} we saw how presence of
excessive missing data can be a smell for incorrect statistical
observations and additional effort for data imputation. Close
attention however must be paid to the distribution of the missing
values within the dataset. Presence of high quantities of missing data
primarily within a column---as opposed to being distributed across
rows and columns---can be a smell that the data is not truly missing.
The missing values in such cases may carry an implicit meaning of a
negative binary response.

This can be validated further by observing the column header along
with the non-missing values of the feature(s) in question. If the
non-missing data is indicative of a positive response such as
\texttt{`\{t,T\}rue' or `\{y,Y\}es'} then the missing data may
indicate a negative response. It is common practise in software
engineering to represent a negative response or result using a null
type, such as \texttt{None} in Python and \texttt{null} in Java. The
same however does not hold in data science as null types are commonly
used to represent missing data.

If a data scientist fails to notice this implicit meaning, they may
hastily drop the missing values or perform imputation. However in
doing so, the original information carried by the dataset is altered
leading to inaccurate results and conclusions.

\subsection{String Value Smells}\label{sec:results-str}
This section presents smells that arise from the presence of string
type data. This group of smells was least frequently observed with a
total occurrence of 12.

\subsubsection{Strings with special characters (str-sanitise)}\label{sec:results-str-sanitise}
The missing values in the \emph{adult} dataset are represented as
\texttt{`?'}. However, the question marks also contain whitespaces,
making their detection and subsequent imputation slightly more
tedious. Another example can be found in the \emph{abalone} dataset.
This dataset contains the \emph{sex} categorical feature where the
value can be one of \texttt{`M', `F' or `I'}. However due to presence
of whitespaces, the data analysis tools may consider values such as
\texttt{` M', `F ', ` I '} valid and distinct from one another.

The presence of leading and trailing whitespaces and special
characters such as punctuation marks in structured data is a smell for
potential problems in the data analysis stage.

Categorical features are often represented as strings during the data
analysis phase and converted to a numerical representation prior to
model training. The presence of whitespace and special characters in
categorical features can confuse data analysis tools and lead to false
results. The problem is easy to rectify in the \emph{abalone} dataset
since the set of correct values is 3. However things become more
challenging for categorical features whose set of values are larger.
For instance, the \emph{airbnb} dataset contains the
\emph{neighbourhood} feature containing names of neighbourhoods in the
United States. The feature contains over 200 valid values and presence
of redundant whitespace and special characters can make the data
cleaning process tedious and time consuming.

Handling presence of special characters in string features requires a
case-by-case analysis and solution. But this smell flags the need to
always check and remove leading and trailing whitespaces from string
features. This is a common task performed by data scientists and
popular data wrangling tools provide built in solutions.

\subsubsection{Numerical features as string (str-num)}\label{sec:results-str-num}
The \emph{playstore} dataset contains data for apps on the Google
Playstore. The dataset contains the \texttt{current\_ver} and
\texttt{android\_ver} features which represent the current version of
the app and the supported android version respectively. The data in
these columns are in the format of release versions such as
\texttt{1.1.9}, which denotes the major, minor and patch versions of
the latest release. Although the information is represented as string,
we can extract 3 separate numerical features here which can provide
valuable insights.

String features may sometimes contain numerical information embedded
within itself. The smell here is the presence of features whose name
indicates numerical type data, but the data analysis tool interprets
the type as string.

Generally speaking, machine learning models tend to perform better
when trained with more data. Extracting valuable numerical information
from such features can therefore be beneficial for model training.

\subsubsection{Strings in human-friendly formats (str-human)}\label{sec:results-str-human}
The \emph{netflix} dataset contains information regarding content on
the popular streaming service. The dataset contains the
\texttt{duration} column which depicts the length of a particular
movie or TV show as depicted in Table~\ref{tab:netflix-duration}. In
the case of movies, the data is clearly numerical in nature (duration
in minutes) but represented as a string format that humans can easily
comprehend (the \emph{str-num} smell discussed earlier in
Section~\ref{sec:results-str-num} also applies here). The data for TV
shows deviates from this format as the duration is represented as the
number of seasons of the TV show. Although this format is also
comprehensible to humans, converting them to a useful numerical
representation however comes with several challenges.

Numerical information being represented in a human-friendly format is
a smell for potential problems during the data analysis stage. This
smell can be considered a subset of the \emph{str-num} smell discussed
earlier in Section~\ref{sec:results-str-num}.

Machine learning models generally perform better when trained with
standardised and uniform data (also see
Section~\ref{sec:results-misc-unit}). In this case, the duration
should be represented in minutes for all examples. Converting the
duration for TV shows from seasons to minutes can be difficult since
the duration of each episode and the number of episodes in a season
may vary amongst TV shows. We may wish to impute using average values
however that requires domain knowledge or further investigation to be
carried out by the data scientist.

\begin{table}[htb]
  \centering
  \caption{Excerpt from the \emph{netflix} dataset showing the
    \emph{str-human} data smell.}
  \begin{tabular}{l l l}
    \toprule
        {} &     type &   duration \\
        \midrule
        0 &    Movie &     90 min \\
        1 &  TV Show &  2 Seasons \\
        2 &  TV Show &   1 Season \\
        3 &  TV Show &   1 Season \\
        4 &  TV Show &  2 Seasons \\
        5 &  TV Show &   1 Season \\
        6 &    Movie &     91 min \\
        7 &    Movie &    125 min \\
        8 &  TV Show &  9 Seasons \\
        9 &    Movie &    104 min \\
        \bottomrule
  \end{tabular}
  \label{tab:netflix-duration}
\end{table}

\section{Discussion}\label{sec:discuss}
This section presents the key observations made in this study.

\subsection{Documentation}\label{sec:discuss-docs}
This study identified several instances where a lack of proper
documentation was felt. The \emph{heart} dataset contains several
cryptic column headers such as \texttt{cp, trestbps, \& fbs}. This
makes it difficult to understand what information the column provides
without prior domain knowledge or further investigation. In the same
dataset, the \texttt{sex} column is label encoded (ie. \emph{male} and
\emph{female} are represented numerically). However without
documentation we cannot ascertain the gender associated with the
numerical values.

Improper documentation makes it difficult to understand the
idiosyncrasies of a dataset such as determining if missing values are
represented with special characters or if they carry an implicit
meaning (see Section~\ref{sec:results-miss}). Developing a deep
understanding of the data is a fundamental step towards any
data-centric work. Documentation can help in this process by providing
useful metadata \& context and help practitioners (re)familiarise
themselves with the dataset. Our observations show the need for tools
that aid machine learning practitioners with documenting their work.
This is also corroborated in prior studies which show that machine
learning practitioners spend considerable amount of time documenting
their work as existing machine learning pipelines, models and datasets
lack proper
documentation~\cite{hutchinson2021towards,kim2017data,sambasivan2021everyone,haakman2021ai}.

In line with recommendations made by
\emph{\citeauthor{hutchinson2021towards}} and
\emph{\citeauthor{sambasivan2021everyone}}, good data documentation
should include (but is not limited to) information regarding the data
source and the data collection procedure, changes that may have
already been made to the dataset along with the rationale behind the
change, expected schema of the columns, meaningful column headers,
presence of missing values \& duplicate rows, correlation amongst
features and descriptive statistics. Providing such documentation can
significantly improve productivity of data scientists and reduce data
understanding and development
time~\cite{hutchinson2021towards,sambasivan2021everyone}.

\subsection{Technical debt}\label{sec:discuss-td}
The analysis revealed technical debt in the datasets due to lack of
best practices and standardised procedures in upstream processes.
Undocumented data practices and transformations
(Section~\ref{sec:discuss-docs}), presence of missing values
(Section~\ref{sec:results-miss}) \& redundant columns
(Section~\ref{sec:results-red}), datasets with sensitive \& imbalanced
columns (Section~\ref{sec:results-misc}) and data in human-friendly
strings (Section~\ref{sec:results-str-human}), all lead to
accumulation of technical debt in downstream stages.

Due to their highly tangled and experimental nature, machine learning
systems are prone to rapid accumulation of technical
debt~\cite{sculley2015hidden,amershi2019software,bosch2021engineering,arpteg2018software}.
Although a holistic view is recommended for monitoring machine
learning systems, it is often difficult to do so due to their
complexity. Data smells however can help detect problems during the
early stages of the machine learning lifecycle when the complexity is
relatively low and fixes are easier and cheaper to implement. As with
traditional software systems, accumulation of technical debt is
inevitable. However early detection can improve effort estimation and
help deliver projects on time with lower financial
costs~\cite{guo2016exploring}.

\subsection{Data validation}\label{sec:discuss-validate}
Data validation tools provide an abstraction over common tests
performed by data scientists when working with data. Analogous to how
regression testing is done when introducing changes to a codebase,
data validation ensures that the new data conforms to certain
expectations when fed into a machine learning
system~\cite{chu2016data}. But writing validation rules still requires
data scientists to understand the data first. This may come naturally
to seasoned practitioners, but is non-trivial for inexperienced
practitioners~\cite{breck2019data}.

As opposed to traditional rule-based software, machine learning models
derive the rules automatically from the data. This reduces the level
of human involvement in such systems, allowing for higher degrees of
automation. Automation however comes at the cost of reduced
transparency as minuscule changes to the input data can cause drastic
changes in the trained model. Data validation and linting tools can
automatically validate the data in terms of correctness, consistency,
completeness and statistical properties. However they lack the ability
to validate dimensions such as fairness and robustness which are
critical in machine learning~\cite{biessmann2021automated}.

We believe that data smells can aid practitioners during the early
stages of data analysis when human involvement is necessary. As seen
in Section~\ref{sec:results-misc-sensitive}
and~\ref{sec:results-cat-hierarchy}, data smells can aid practitioners
to catch data quality issues that lead to biased and unfair
predictions in their models. The smells also help fix other data
quality issues that results in a more robust dataset.

\subsection{Data Efficiency}\label{sec:discuss-efficient}
The availability of highly resilient and cheap hardware commodity due
to cloud-computing has enabled leaping advancements in AI. Neural
networks have consistently evolved in complexity and size, starting
with Imagenet in 2009, Resnet in 2015 and more recently GPT-3 in 2020
which consists of 175 billion parameters~\cite{brown2020language}.
Complex models are data hungry, requiring training data in the scale
of Terabytes. In this era of big data and high performance computing,
presence of seemingly minor data smells can lead to wasted training
cycles and cost millions~\cite{dhar2020carbon}. Engineering efficient
datasets become crucial in such circumstances and circumvent the use
of overly complex machine learning models. Data quality plays a key
role in delivering efficient and maintainable models with a smaller
carbon footprint.

\section{Threats to Validity}\label{sec:limit}
This study opted for a \textit{shallow} analysis of the datasets. The
analysis phase consisted of a specific set of steps that were carried
out (as outlined in Section \ref{sec:method}) for each dataset. This
was a deliberate decision, made to easily scale and reproduce the
analysis steps across a large sample of datasets. We recognise that a
\textit{deeper} analysis of each dataset may reveal further smells.
For instance, the current analysis excluded \textit{outlier detection}
or fitting machine learning models to the dataset. However we strongly
believe that the generalisability of smells reduces as we increase the
depth of the analysis. That is, the smells would only be valid within
the context of the problem domain.

The smells discovered by this study are linked to the version of the
data used for the analysis. For instance, \textit{Kaggle} contains
datasets from the \textit{UCI Machine Learning Data Repository
  (UCI)}\footnote{https://archive.ics.uci.edu/ml/datasets.php}.
However the version of data found on Kaggle is different from the
original source on UCI. Similarly, it is also unclear if the version
of data hosted on UCI is the ground truth or was derived from
somewhere else. While we recognise this issue, it is also
unfortunately the nature of all data science problems.
Table~\ref{tab:datasets} provides a list of all datasets that were
used in this analysis along with their version at the time it was
downloaded in hopes to increase the reproducibility of the results.

The current analysis does not extend to quantifying impact of smells.
For instance, we do not know if and to what extent the \textit{unknown
  unit of measure (misc-unit)} smell affects the model's performance
on a test set (see Section~\ref{sec:results-misc-unit}). Such a task
requires collection of a sufficiently large sample of datasets that
contain the smell in question. The validation process involves
performing a supervised learning task which is a highly experimental
and time consuming process. The effort and time increases when we
scale the validation across multiple smells. Therefore such a time
consuming task was beyond the scope of this study.

Data smells are subject to the interpretation of the data scientist,
the team or the organisation performing the analysis. This problem of
subjectiveness is present in code smells as well. Not all long methods
are bad and god classes still exist in open source repositories. To
reduce the possibility of subjective bias in our results, the smells
were reviewed by the second author prior to including them in the
catalogue.

\section{Conclusion}\label{sec:conclude}
Code smells are frequently used by software engineers to identify
potential bugs, sources of technical debt and weak design choices.
Code smells in the context of traditional software have existed for
over three decades and have been extensively studied by the software
engineering research community. With the growing popularity of AI and
its adoption in high-stakes domains where a data-centric approach is
adopted, data smells are seen as a much needed aid to machine learning
practitioners. This study examined 25 public datasets and identified
14 recurrent data quality issues---coined as data smells---that can
lead to problems when training machine learning models. Our results
indicate a need for better data documentation, and accumulation of
technical debt due to lack of standardised practices in upstream
stages of machine learning pipelines.

We consider our collection of data smells and the analysis of their
prevalence a first step towards aiding data scientists in the initial
stages of data analysis where human involvement is necessary. We hope
that our work raises awareness amongst practitioners to write better
documentation for their datasets and follow best practises during data
collection to minimise technical debt in upstream stages. As a next
step, we aim to grow the data smells catalogue by analysing more
datasets. Furthermore, we wish to remove the constraints introduced by
\emph{IC2} (see Section~\ref{sec:method} and
Table~\ref{tab:dataset-ic}) and include datasets larger than 1GB in
size.

\bibliographystyle{ACM-Reference-Format}
\bibliography{report.bib}
\end{document}